\documentclass[reprint,secnumarabic,amsmath,amssymb, nobibnotes, aps, prl,superscriptaddress]{revtex4-2}

\usepackage{graphicx}
\usepackage{dcolumn}
\usepackage{bm}
\usepackage[locale=US,separate-uncertainty=true]{siunitx}
\usepackage[english]{babel}
\usepackage[version=4]{mhchem}
\usepackage{float}
\usepackage{braket}
\usepackage{dsfont}
\usepackage{microtype}
\usepackage{mathtools}
\usepackage{hyperref}

\newcommand{\thefontsize}{The current font size is: \f@size pt}

\newcommand\mymapsto{\mathrel{\ooalign{$\rightarrow$\cr%
  \kern-.15ex\raise.275ex\hbox{\scalebox{1}[0.522]{$\mid$}}\cr}}}

\begin{document}

\title{Expressivity of Quantum Reservoir Computers}

\author{Nils-Erik Schütte}
\affiliation{Carl von Ossietzky Universität Oldenburg, Fakultät V, Institut für Physik, 26129 Oldenburg, Germany}
\affiliation{DLR, Institute for Satellite Geodesy and Inertial Sensing, Am Fallturm 9, 28359 Bremen, Germany}

\author{Niclas Götting}
\affiliation{Carl von Ossietzky Universität Oldenburg, Fakultät V, Institut für Physik, 26129 Oldenburg, Germany}

\author{Hauke Müntinga}
\affiliation{DLR, Institute for Satellite Geodesy and Inertial Sensing, Am Fallturm 9, 28359 Bremen, Germany}

\author{Meike List}
\affiliation{DLR, Institute for Satellite Geodesy and Inertial Sensing, Am Fallturm 9, 28359 Bremen, Germany}
\affiliation{University of Bremen, 28359 Bremen, Germany}

\author{Daniel Brunner}
\affiliation{Institut FEMTO-ST,  Universit\'e Franche-Comt\'e CNRS UMR 6174, Besan\c{c}on, France}

\author{Christopher Gies}
\affiliation{Carl von Ossietzky Universität Oldenburg, Fakultät V, Institut für Physik, 26129 Oldenburg, Germany}


\begin{abstract}
Using Hamiltonian encoding to inject an input into parameterized quantum circuits (PQCs), the output of the PQC can be written as truncated Fourier series.
In recent years, the expressivity of PQCs was established as the number of frequencies contained in this Fourier series.
While this concept has also been applied to other quantum machine learning (QML) paradigms, a clear notion of expressivity for temporal information processing with quantum systems is still lacking.
Here, we introduce such a notion to the field of quantum reservoir computing (QRC).
We analytically derive an expression for the readouts showing that the output of a QRC can be interpreted as a multi-dimensional Fourier series.
We give a formula for the growth of expressivity induced by the sequential information injection, which we corroborate with numerical simulations, calculating explicitly the number of multi-dimensional output functions which can be generated from the readouts.
Our results show that the specific interplay between system size, input encoding,  and memory time gives rise to a boundary on the system size beyond which it is obstructive to further increase the reservoir size in extreme scrambling systems.
We propose a recipe for determining this maximal system size for a given QRC setup.

\end{abstract}

\maketitle


\section{Introduction \label{sec_introduction}}

The pursuit of quantum advantage in information processing has motivated the development of quantum machine learning (QML) algorithms and methods, leveraging quantum circuits or dynamical quantum systems to offer potentially exponential speedup or capacity over classical methods. Interest in QML has further increased as meaningful computation involving quantum circuits became possible on existing quantum hardware.
In contrast to well recognized algorithms like Shor and Grover \cite{shor1994algorithms, shor1997polynomialtime, grover1996fast}, the so-called ansatz in QML requires few gate operations only, minimizing gate-fidelity and qubit coherence-time requirements imposed by algorithmic quantum computing.
The gate-based approach to QML leverages parameterized circuits that are adjusted to minimize a cost function, which can be viewed as iterative topology improvements of the quantum computing circuit  \cite{mitarai2018quantum, schuld2019quantum, havlicek2019supervised}.
We refer to this approach as parametrized quantum circuit quantum machine learning (PQC-QML).
Seemingly unrelated, another QML approach called quantum reservoir computing (QRC) has emerged from the origin of artificial neural networks that harnesses the inherent dynamics of quantum systems for predicting non-linear functions of input data.
In reservoir computing, the interconnects between the input and the neurons as well as among the neurons themselves are not trained but remain fixed, and learning is performed only at the generation of the output in a linear layer.
This extends the possible substrates beyond quantum circuits to a variety of open quantum systems, photonic platforms and other implementations, and some of them have been already realized \cite{carles2026experimental, paparelle2026experimental}.
Belonging to the class of recurrent neural networks, QRC is being considered especially for processing time-series data \cite{fujii2017harnessing, cindrak2024enhancing, dudas2023quantum, mujal2023timeseries}, albeit classification has been demonstrated as well \cite{suzuki2022natural}.
The key motivation is to harvest the exponential scaling in Hilbert-space dimensionality, as compared to a linear relationship between dimensionality and number of hardware neurons in classical physical hardware systems.

QRC has been around for a much shorter time than PQC-QML, and its predictive capabilities have not yet been explored in the same depth.
Over the past years, QRC performance has been gauged using measures based on dimensionality \cite{govia2021quantum, llodra2023benchmarking, gotting2023exploring, cindrak2024krylov,kalfus2022hilbert}, short-term memory capacity \cite{jaeger2001short, kobayashi2024feedbackdriven, sannia2024dissipation} as well as a variety of non-linear benchmark tasks \cite{fujii2017harnessing, cindrak2024enhancing, settino2025memoryaugmented}.
Furthermore, benefits as well as the usual physical substrate-induced challenges faced when leveraging physical quantum effects for computing have been investigated in the QRC context, studying the impact of dissipation \cite{fujii2017harnessing, olivera-atencio2023Benefits, sannia2024dissipation, kora2024frequency, gotting2025connection, domingo2023taking}, entanglement \cite{gotting2023exploring}, quantum phases \cite{martinez-pena2021dynamical, llodra2025quantum, kobayashi2026edge} and coherence \cite{palacios2024role}.
These studies identified correlations between the reservoir's quantum properties and fundamental performance characteristics.
However, a clear notion of expressiveness, as it is known in PQC-QML \cite{schuld2021effect} and which has recently been extended to quantum extreme learning machines \cite{xiong2025fundamental}, is still lacking.

In this work, we introduce a notion of expressivity to the field of temporal information processing with quantum systems by analytically showing that the output of a QRC can be written as a multi-dimensional Fourier series where each dimension originates from a particular input time step.
We quantify expressiveness by the number of linearly independent output functions which can be generated from the readouts and give explicit formulas for the increasing expressivity with the QRC cycles for different input encoding strategies.
We support our analytical findings with numerical simulations using the framework of the resolvable expressive capacity (REC) and eigentasks \cite{hu2023tackling} and apply it to a Haar random reservoir.

Recent studies showed that the scaling of QRC systems is limited due to the exponential concentration of the outputs \cite{xiong2025role, sannia2025exponential},  effectively restricting the time for which information can be stored when increasing the system size.
Together with the fading memory property, this gives rise to a maximal size of the reservoir beyond which it is superfluous to add more qubits.
Consequently, together with the fading‑memory property, a maximal reservoir size emerges beyond which adding more qubits becomes superfluous.
In this work we provide an explicit description of how to determine that size.

The paper is organized as follows.
In Sec. \ref{sec:quantum_reservoir_computing_setup} we introduce our QRC setup including the input injection and the readout.
In Sec. \ref{sec:multivariate_fourier_decomposition} the established notion of expressivity is extended to time-series problems from the field of temporal information processing with quantum systems.
In particular, we give a formula for the theoretical possible growth of expressivity with QRC cycles under idealized assumptions.
Sec. \ref{sec:expressivity_in_quantum_reservoir_computing} contains our central findings on expressivity of quantum reservoir computers, for which we investigate its dependence on the system size and the input encoding. It turns out that a particular focus must be placed on the interplay of memory time, system size, and the input encoding scheme. From the insight that is obtained, it is then possible to formulate a recipe on how to determine an optimal reservoir size avoiding unnecessary memory restriction.

\section{Quantum reservoir computing setup} \label{sec:quantum_reservoir_computing_setup}

\begin{figure}
    \centering
    \includegraphics[width=\linewidth]{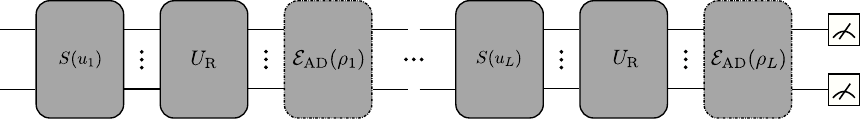}
    \caption{
        Schematic figure illustrating the implementation of a QRC scheme on a gate-based quantum computer.
        We assume that the qubits are initialized in a random state.
        $\mathcal{E}_{\text{AD}}$ denotes the applied amplitude damping channel according to Eq. \eqref{eq:krauss_channel}.
    }
    \label{fig:qrc_circuit_schematic}
\end{figure}

Generally, we consider a QRC setup as depicted in Fig. \ref{fig:qrc_circuit_schematic}, with $Q$ qubits and the following components:
\begin{enumerate}
    \item Input encoding:
    We consider a discrete time input signal $u_l \in [0, 2 \pi]$.
    To inject this data into the reservoir, we use a gate $S(u_l)$, composed of single-qubit rotations, as these are the most commonly employed input mechanism in current QRC implementations \cite{sannia2024dissipation,mifune2025effects,tovey2025generating}.
    Instead of the usual $\sigma_i, i \in [x, y, z]$ encoding, we apply a more general rotation encoding around randomly sampled axes on a tunable number of input qubits $Q_{\text{in}}$.
    The input gates are then of the form $\text{e}^{-\text{i} \beta_{q_{\text{in}}} u G}$, with $q_{\text{in}} = 0, \ldots, Q_{\text{in}}$ and the encoding Hamiltonian $G$ being a summation of Pauli matrices:
\begin{align}
    G = \frac{1}{2} (n_x \sigma_x + n_y \sigma_y + n_z \sigma_z).
\end{align}
The $n_{i \in [x, y, z]}$ are randomly sampled for each qubit individually such that they fulfill $n_x^2 + n_y^2 + n_z^2 = 1$.
The corresponding unitary gate can be decomposed as
\begin{align}
    \begin{aligned} \label{eq:rotation_around_arbitrary_axis}
    R_{n}(u) &= \text{e}^{-\text{i} \beta_{q_{\text{in}}} u G} \\
    &= R_z(\alpha) R_y(\beta) R_z(\beta_{q_{\text{in}}} u) R_y(- \beta) R_z(- \alpha) ,
    \end{aligned}
\end{align}
with
\begin{align}
    \begin{aligned}
    \alpha &= \text{arctan2}(n_y, n_x) , \\
    \beta &= \arccos(n_z) . \nonumber
    \end{aligned}
\end{align}
Describing the state of the system at time step $l$ by a density matrix $\rho_l \in \mathbb{C}^{D \times D}$, where $D = 2^Q$ is the Hilbert space dimension, the input-injection process can be written as
\begin{align}
    \rho_l \mapsto S(u_l) \rho_l S^\dagger (u_l) .
\end{align}
A random pure state $\rho_0$ serves as our initial state, but in our numerical simulations we apply a washout sequence of 100 time steps before any performance analysis.

Throughout this work, we consider two types of input encoding: (a) linear encoding with $\beta_{q_{\text{in}}} = 1, \, \forall q_{\text{in}}$, and (b) exponential encoding introduced in Ref. \cite{shin2023exponential} with $\beta_{q_{\text{in}}} = 3^{q_{\text{in}}}, \, q_{\text{in}} = 1, \ldots, Q_{\text{in}}$.
    \item Reservoir dynamics:
    Most QRC implementations focus on specific physical dynamics, a common choice being the transverse-field Ising model (TFIM) \cite{fujii2017harnessing,gotting2023exploring, cindrak2024enhancing}.
    The reservoir time evolution is then directly determined by the unitary operator $U_{\Delta t}$ obtained from the model Hamiltonian $H$ as
\begin{align}
    & U_{\Delta t} = \text{e}^{- \text{i} H \Delta t} \label{eq_time_evolution_operator} ,\\
    & \rho \mymapsto U^{\phantom \dagger}_{\Delta t} \rho U_{\Delta t}^\dagger \label{eq:dm_time_evolution} ,
\end{align}
While the Hamiltonian approach offers physical interpretability, it also introduces the problem of finding optimal clock cycles $\Delta t$ to facilitate sufficient information spreading for computation \cite{cindrak2024krylov}.
As we are interested in fundamental, upper limits of the expressiveness of QRC, instead of choosing a specific physical system, we draw a random unitary $U_{\text{R}}$ from the Haar distribution, such that $U_{\Delta t} \equiv U_{\text{R}}$ \cite{mezzadri2007how}.
Haar random unitaries maximally scramble information \cite{holmes2021barren, roberts2017chaos}, allowing us to avoid any restriction on the quantum dynamics that may arise from a particular choice of a Hamiltonian.
To ensure the fading memory property \cite{jaeger2001short} of the QRC model, we use an amplitude damping channel to simulate qubit decay.
Every qubit is subject to a damping channel \cite{nielsen2010quantum}
\begin{align}
    & \rho \mapsto \sum_i K_i \rho K_i^\dagger , \label{eq:krauss_channel} \\
    & K_0 = \begin{pmatrix}
        1 & 0\\
        0 & \sqrt{1 - p}
    \end{pmatrix}, \quad K_1 = \begin{pmatrix}
        0 & \sqrt{p}\\
        0 & 0
    \end{pmatrix} ,
\end{align}
with decay probability $p$.
It has the following relationship to the decay rate $\gamma$:
\begin{align}
    p = 1 - \text{e}^{-\gamma t} .
\end{align}
    \item Readouts:
    In the most general case, information extraction from a qubit system can be described as measuring the expectation values of $P \leq D^2$ elements $\{O^{(p)}\}_{p \in [|0, P-1|]}$ of a positive operator-valued measure (POVM).
    Generally, $P$ can be larger than the Hilbert space dimension $D$, but we set $P = D$ unless specified otherwise.
    We choose the $p$-th POVM element $M_p$ to be a zero matrix with 1 at the $p$-th position on the diagonal ($M_p = \ket{p}\bra{p}$), effectively measuring  all diagonal elements of the density matrix.
    Hence, readouts at time step $l$ are given by the expectation values of the operators $O^{(p)} = M_p^\dagger M_p$:
\begin{align}
    \langle O^{(p)} \rangle = \text{Tr} \{O^{(p)} \rho_l \}, \quad p = 0, \ldots, P - 1 .
\end{align}
    \item Classical post-processing:
    The measured observables are linearly combined as
    \begin{align} \label{eq:prediciton_linear_combination}
        f_\eta (\bm{u}) = \sum_{p = 0}^{P - 1} \eta_p \langle O^{(p)} \rangle ,
    \end{align}
    where the  weights $\bm{\eta} = (\eta_0, \ldots, \eta_{P - 1})^{\text{T}}$ are trained classically via linear regression.
\end{enumerate}
As was pointed out before, the time evolution in Eq.~\eqref{eq:dm_time_evolution} natively translates into the language of quantum circuits, allowing the unitary operation to be realized as a circuit on a gate-based quantum computer \cite{fujii2017harnessing, hamhoum2025multivariate, hu2024overcoming}.
%
We explicitly exploit this connection in Sec. \ref{sec:multivariate_fourier_decomposition} to give an upper bound for the expressivity of the QRC framework.

\section{Multivariate Fourier decomposition} \label{sec:multivariate_fourier_decomposition}

The expressive power in QRC determines how well a quantum reservoir is able to approximate any given target function.
We here employ a similar method for expressivity analysis as established by Schuld et al. \cite{schuld2021effect} for PQC-QML and later applied to quantum extreme learning machines (QELMs) \cite{xiong2025fundamental}.
In particular, we quantify the expressivity via the number of linearly independent functions the quantum reservoir can generate.

While the above QML frameworks are feed forward type machine learning approaches, QRC is an intrinsically temporal information processing paradigm, where inputs are injected into the system sequentially.
As introduced in Section~\ref{sec:quantum_reservoir_computing_setup}, QRC additionally features a dissipative process to ensure fading memory.
These two peculiarities set the expressivity analysis in QRC apart from existing endeavors in the field of QML.
Importantly, the dissipative dynamics poses a challenge to exact analytical derivations of QRC expressivity, as it depends on present noise channels and their specific influence on system dynamics, prohibiting a concise discussion.
We will, thus, at this point take a step back from the conventional QRC framework by excluding dissipative effects from the derivation, leading to an upper bound of expressivity in QRC.

%
By treating the quantum reservoir computer as a sequence of input and time evolution gates repeated $L$ times, we can infer the expressive power from the number of Fourier components present in the reservoir state after the last gate is executed.
For the derivation, we transform the dynamics such that the input gate is diagonal and of the form $S(u_l) = \sum_{d = 0}^{D - 1} \text{e}^{- \text{i} \lambda_d u_l} \ket{d} \bra{d}$, with $\lambda_d$ being the eigenvalues of the generating Hamiltonian $G$, $u_l$ the input at time step $l$, and $d$ enumerating the basis states.

After initializing the system in a state $\rho_0 = \sum_{i,j = 0}^{D - 1} \rho_{ij} \ket{i} \bra{j}$ a random washout sequence of 100 time steps is applied to erase the dependence on this initial state.
Subsequently the input encoding gates $S(u_l)$ and the unitary reservoir dynamics $U_{\text{R}} = \sum_{i,j = 0}^{D - 1} w_{ij} \ket{i} \bra{j}$ are successively applied.
The expectation value of observable $O^{(p)}$ is then given by
\begin{align}
    \langle O^{(p)} \rangle (\bm{u}) & = \text{Tr} \{ O^{(p)} \rho \} \nonumber \\
    & = \sum_{\bm{\omega} \in \Omega} \text{e}^{\text{i} \bm{\omega} \bm{u}} c_{\bm{\omega}} , \label{eq:observables_fourier_series}
\end{align}
with the spectrum of vectorial Fourier frequencies
\begin{align} \label{eq:frequency_spectrum_0}
    \Omega = \{\bm{\lambda}_{\bm{d}^\prime} - \bm{\lambda}_{\bm{d}} \, | \,(\bm{d}, \bm{d}^\prime) \in [|0, D-1|]^{\times L} \}
\end{align}
and their respective Fourier components $c_{\bm{\omega}}$.
In Eq. \eqref{eq:observables_fourier_series} we have collected the input series up to input $L$ into a vector $\bm{u} = (u_1, \ldots, u_L)^{\text{T}}$ that naturally arises in the derivation as is spelled out in details in Appendix~\ref{app:fourier_series_representation_of_qrc}.
This result enables us to specify the action of a quantum reservoir on a univariate input time series in terms of a multivariate truncated Fourier series.
To support this view, one may conceptually consider time-series processing via a static QELM or even a PQC approach by injecting the complete time series up to time step $L$ at once into the system, which would likewise result in a multivariate output.

We now assess the number of linearly independent functions that are contained in a single expectation value, by determining the number of distinct frequency vectors in the frequency spectrum $\Omega$, in which each entry arises from a specific input injection step.
Since the same input gate is applied at every time step, each component of $\bm{\omega}$ can take on the same values, given by the possible differences of the eigenvalues $\lambda_d$ of the encoding Hamiltonian.
With this, we can rewrite Eq.~\eqref{eq:frequency_spectrum_0}:
\begin{align}
    \Omega_0 = \{ \omega = \lambda_{d^\prime} - \lambda_d \, | \, d, d^\prime \in [|0, D - 1|] \} \\
    \Omega = \{ \bm{\omega} = (\omega_0, \ldots, \omega_{L - 1})^{\text{T}} \, | \, \omega_l \in \Omega_0 \} .
\end{align}

For $\Omega_0$, this corresponds to the set of Fourier frequencies for a QELM, for which the maximal possible number of distinct eigenvalue differences was derived in Ref. \cite{xiong2025fundamental}: Assuming a general encoding gate $S(u_l)$ acting on $Q_{\text{in}}$ input qubits with distinct eigenvalues $\lambda_0, \ldots, \lambda_{D_{\text{in}} - 1}$ ($D_{\text{in}} = 2^{Q_{\text{in}}}$), the number of possible eigenvalue differences (and, hence, non-vectorial frequencies) is given by $D_{\text{in}} (D_{\text{in}} - 1) + 1 = 4^{Q_{\text{in}}} - 2^{Q_{\text{in}}} + 1$.
As each subsequent layer combines these Fourier components with all previous functions, the number of distinct frequencies scales exponentially with the layer depth $L$ -- independently of the specific input encoding.
This determines the maximal number of distinct vectorial frequencies in $\Omega$ after $L$ steps to be $\left( 4^{Q_{\text{in}}} - 2^{Q_{\text{in}}} + 1 \right)^L$.
The linear and exponential input encodings considered in this work impose restrictions on the eigenvalues of $S$, thus reducing the expressivity to
\begin{align}
    & \text{linear:} \quad && (2 Q_{\text{in}} + 1)^L \label{eq:frequency_scaling_lin} \\
    & \text{exponential:} && (3^{Q_{\text{in}}})^L = 3^{Q_{\text{in}} L} \label{eq:frequency_scaling_exp} ,
\end{align}
where the individual scaling behaviors for a single time step were derived in Refs. \cite{schuld2021effect, shin2023exponential}.

%
We note at this point that the expressivity only depends on the number of layers and input qubits, \textit{not} the reservoir size $Q$.
This is in stark contrast to classical RC, where the available output function space grows upon increasing the size of the underlying system.
The linearity of quantum mechanics entirely precludes the emergence of more orthogonal functions via the system dynamics, leaving only the input encoding as a source of expressivity.
While appearing as a strong limitation of QRC at first sight, the exponential scaling with respect to layer depth in combination with the fact that RC in general deals with large time series counteracts this drastically, effectively eliminating the limitation for any meaningful learning task.

The reservoir size instead acts as an upper bound for the attainable expressivity in QRC, analogously to classical RC.
A general mixed quantum state of $Q$ qubits possesses $4^Q - 1$ degrees of freedom, thus allowing for an absolute maximum of $4^Q - 1$ linearly independent functions of the input.
In most realistic scenarios, however, not all degrees of freedom in the quantum system are measurable, further reducing the expressivity to the $P$ readout functions, which in this work are composed of the $2^Q$ diagonal elements of the density matrix.
An effective means of increasing the readout dimension is time multiplexing \cite{fujii2017harnessing}.
In introducing $V$ virtual nodes to the reservoir computer via subsampling of the system dynamics, it increases the available function space to $V\cdot P$, while naturally still being upper bounded by the total system size.
These limits combined with the expressivity scaling behavior lead to a recipe for optimal reservoir sizes, which we explore in-depth in Sec.~\ref{sec:optimal_reservoir_size}.

All analyses up to this point were only concerned with theoretical upper bounds of expressivity in QRC, disregarding the fading memory property imposed by dissipative dynamics.
Additionally, due to QRCs stochastic nature, experiments on quantum reservoirs are fundamentally subject to shot noise.
These two effects surpass the capabilities of the purely analytical approach in this chapter, but represent intrinsic characteristics of any QRC implementation, thus demanding for a deeper look into how they affect the learning performance in QRC.
In the next section, we will therefore apply a more sophisticated measure that includes all of the above effects.

\section{Expressivity in quantum reservoir computing} \label{sec:expressivity_in_quantum_reservoir_computing}

We support and extend our theoretical analysis from Sec. \ref{sec:multivariate_fourier_decomposition} with numerical simulations by employing the concept of resolvable expressive capacity (REC) \cite{hu2023tackling}.
Before incorporating the experimental reality of shot noise into the analysis, we first lay out the general idea behind the REC in the following.

Within the REC framework, the quantum system is considered a black box input-output map that generates features for learning; the mechanisms behind said map here being the same as in the above sections.
Let $\bm{x}(\bm{u}) = \left( x_0(\bm{u}), \ldots, x_{R - 1}(\bm{u}) \right)^{\text{T}}$ denote the exact POVM measurement expectation values after injecting the inputs $\bm{u}$, then we obtain the Gram matrix $\bm{G}$ and covariance matrix $\bm{V}$ of the quantum feature map as
\begin{align}
    \bm{G} &= \mathbb{E}_{\bm{u}} [\bm{x} \bm{x}^{\text{T}}] \nonumber \\
    \bm{V} &= \bm{G} - \bm{D} \nonumber \\
    D_{rr} &= \mathbb{E}_{\bm{u}} [x_r] \nonumber ,
\end{align}
where $\mathbb{E}_{\bm{u}}$ is the expectation value over the input domain and $\bm{D}$ is a diagonal matrix.

To gather an understanding about the richness and structure of the feature map, we solve the generalized eigenvalue problem
\begin{align}
    \bm{V} \bm{r}^{(r)} = \beta_r^2 \bm{G} \bm{r}^{(r)}
    \label{eq:rec_eig_prob}
\end{align}
for the noise-to-signal ratios $\beta^2_r$ and eigenvectors $\bm{r}^{(r)}$.
The eigenvectors are functions in the space of measured features which define a linear transformation
\begin{align}
    y^{(r)}(\bm{u}) = \sum_j r_j^{(r)} x_j (\bm{u}),
\end{align}
yielding the so-called eigentasks, which are native to the physical system in the sense that they identify the least noisy functions the system can generate, each associated with its respective noise-to-signal ratio $\beta^2_r$.

The eigenvalue problem of Eq.~\eqref{eq:rec_eig_prob} so far assumes infinite measurement shots and an exact expectation value over all possible input strings, the former being experimentally unachievable and the latter even numerically.
Hence, we will at this point introduce corrections to the exact theory.
Instead of extracting exact features $\bm{x}$ from the quantum system, we approximate the readouts via $S$ measurements of single-shot features $X_{r}^{(s)}(\bm{u})$ that depend on their respective input string, obtaining the averaged readouts
\begin{align} \label{eq:average_stochastic_features}
  \bar{X}_r (\bm{u}) = \frac{1}{S} \sum_{s=1}^{S} X_{r}^{(s)} (\bm{u}) .
\end{align}
We then average these over $\tilde{N} \coloneqq N^{\tilde{l}}$ different input sequences $\bm{u}$, where we choose $N=20$ inputs per layer for a system with $\tilde{l} = 3$ layers, resulting in a total $8000$ distinct input series.
These statistics yield an approximate version of the eigenvalue problem of Eq.~\eqref{eq:rec_eig_prob}, from which we obtain noise-to-signal ratios $\beta^2_{\tilde{N},r}$ that, however, strictly underestimate the exact values.
As detailed in \cite{hu2023tackling}, the correction
\begin{equation}
  \beta^2_r \approx \frac{S \beta^2_{\tilde{N},r}}{(S - 1) - \beta^2_{\tilde{N},r}}
\end{equation}
allows to infer the noise-to-signal ratios of the exact equations from the finite statistics values.

We finally employ the aforementioned modifications to obtain the approximate REC of the quantum system as
\begin{align}
    C_{\text{T}} = \sum_{r \in \tilde{R}} \frac{1}{1 + \frac{\beta_r^2}{S}} .
\end{align}
Importantly, the sum does not run over all POVMs, but only over the subset of linearly independent features $\tilde{R}$, which may be reduced due to symmetries in the reservoir.
The value $C_{\text{T}}$ quantifies the capacity of the system to approximate its eigentasks, i.e., it not only determines the number of linearly independent output functions the system can express, but also how well so.
The REC thus generalizes the information processing capacity \cite{dambre2012information} by natively incorporating the effects of sampling noise \cite{polloreno2025restrictions} into the framework.
We also note at this point that no further modifications have to be made to include dissipative effects into the theory, due to the abstract black box approach of the REC.

\begin{figure}
    \raisebox{-0.5\height}{\includegraphics[width=0.48\linewidth]{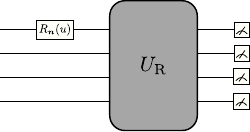}} \hfill
    \raisebox{-0.5\height}{\includegraphics[width=0.48\linewidth]{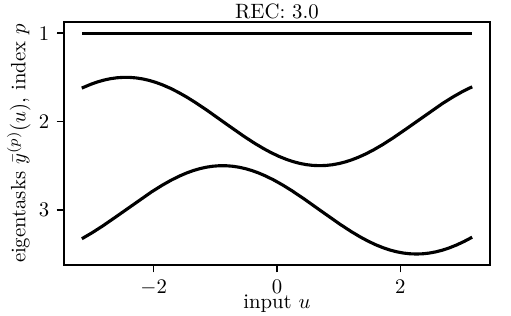}}
    \caption{Left: Quantum circuit implementing one time step of the reservoir dynamics using one encoding gate and a Haar random unitary.
    Right: The corresponding eigentasks of the system.
    The eigentasks are scaled for better visibility.}
    \label{fig:circuit_1_encodings_new}
\end{figure}

\begin{figure}
    \raisebox{-0.5\height}{\includegraphics[width=\linewidth]{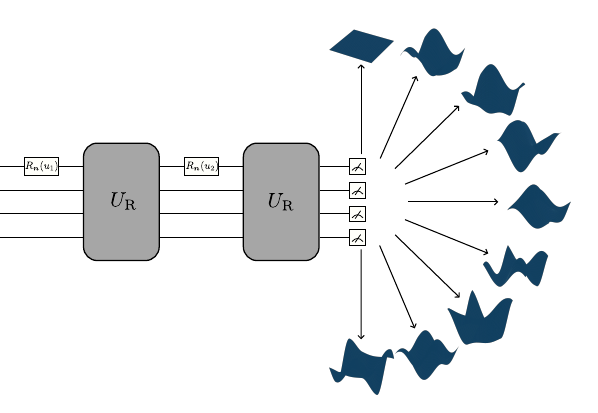}}
    \caption{Schematic figure illustrating the construction of two-dimensional output functions when using one rotational input gate and two time steps in QRC.
    The reservoir dynamics are realized by a Haar random unitary.}
    \label{fig:eigentask_example_2d}
\end{figure}

One of our contributions is to adopt the REC analysis to the field of temporal information processing by calculating multi-dimensional eigentasks.
The way we extend the existing REC analysis to a sequential input is detailed in App. \ref{app:multivariate_rec_analysis}.
We first apply the REC analysis to a quantum reservoir with a single input gate as depicted in the Figures \ref{fig:circuit_1_encodings_new} and \ref{fig:eigentask_example_2d}.
The system consists of four qubits that are initialized in a random state.
Input is injected via a single-qubit Pauli rotation, in this case around a randomly sampled axis according to Eq. \eqref{eq:rotation_around_arbitrary_axis}.
Since the reservoir's Haar random unitary does maximally scramble information across the different degrees of freedom \cite{holmes2021barren, roberts2017chaos}, we avoid any restriction on the expressivity coming from the particular design of the underlying reservoir.
For this first example, we neglect measurement statistics (infinite shots) and extract results of the POVM directly from the density matrix.

The right panel of Fig. \ref{fig:circuit_1_encodings_new} shows the three eigentasks we obtain from employing the REC analysis for infinite shots on these measurements: a sine, a cosine, and a constant after the first reservoir cycle.
Going through another QRC cycle, the possible output function space is now two-dimensional with 9 orthogonal output functions, basically given by all two-dimensional combinations of the three output functions after the first cycle, as illustrated in Fig. \ref{fig:eigentask_example_2d}.
These two-dimensional eigentasks were calculated with the multi-dimensional eigentask analysis which we detail in App. \ref{app:multivariate_rec_analysis}.
It should be noted that the eigentasks generally do not coincide with the basis functions that one would naturally choose when dealing with Fourier series.
Instead, they are specifically the orthogonal functions that can be approximated by the system with minimal error.
Due to the infinite shot scenario in these explanatory examples, the $C_{\text{T}}$ corresponds exactly to the rank of the Gram matrix, as the influence of the noise-to-signal ratios vanishes in the limit of exact measurements.
In the following sections, we thus investigate the finite-shot case and generalize our results to a variable number of reservoir qubits and more than two time steps.

\subsection{Reservoir size and memory time} \label{sec:reservoir_size_and_memory_time}

While the multivariate REC analysis can be done, in principal, for any number of variables, we restrict our numerical simulations to $\tilde{l} = 3$ most of the time, due to the exponentially scaling computational cost emerging from the input string sampling.
We, however, do include further time steps beyond the sampled input strings to illustrate the fading memory via the dissipative channel of the quantum reservoir, as detailed in App. \ref{app:multivariate_rec_analysis}.
To verify upper bounds of the REC analysis also for the larger quantum reservoirs in this section, we again first determine $C_{\text{T}}$ for the infinite shot scenario to then proceed to a reservoir, whose readout precision is limited by sampling noise.

\begin{figure}[h]
    \centering
    \includegraphics[scale=1]{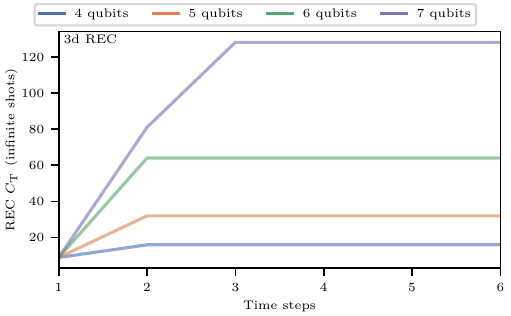}
    \caption{Three dimensional resolvable expressive capacity in dependence on the QRC cycles.
    The input is injected with the linear encoding scheme on $Q_{\text{in}}=4$ input qubits.
    The reservoir dynamics are realized by a Haar random unitary and the results are averaged over ten realizations.
    The REC has been calculated with exact expectation values, using effectively $S \rightarrow \infty$ measurements.
    Shaded areas show the standard deviation.
    The small standard deviations indicate that different random unitaries drawn from the Haar measure do nearly equally scramble information.
    For the first two time steps, the eigentasks are, of course, not three dimensional but one and two dimensional, respectively.
    For the simulation a decay rate of $\gamma = 0.04$ has been used.}
    \label{fig:rec_time_steps_infinite_3_0.04}
\end{figure}
Fig. \ref{fig:rec_time_steps_infinite_3_0.04} shows the resolvable expressive capacity produced by reservoirs of different sizes $Q \in [|4, 7|]$, a constant four input qubits and an amplitude damping rate $\gamma = 0.04$.
At the first time step, we can see that each system resembles the expressivity of a QELM with $C_{\text{T}} = 2 Q_{\text{in}} + 1 = 9$.
The REC then continues to grow according to the analytic derivations of Sec.~\ref{sec:multivariate_fourier_decomposition}, until limited by the $2^Q$ POVMs that provide an absolute upper bound for the RECs.
We here note that the effect of dissipative dynamics on the REC vanishes in the infinite shot scenario, as the arbitrarily precise measurements allow definite feature extraction even at the smallest scales.

\begin{figure}[h]
    \centering
    \includegraphics[scale=1]{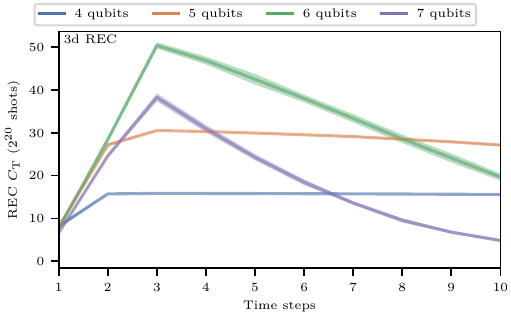}
    \caption{
        Same as Fig. \ref{fig:rec_time_steps_infinite_3_0.04}, but for $S = 2^{20}$ measurement shots.
        REC values are also averaged over ten realizations.
    }
    \label{fig:rec_time_steps_finite_lin_3_20_0.04}
\end{figure}
In real experiments it is, however, impossible to obtain exact expectation values and the REC framework proves to be a powerful means of assessing, whether theoretical performance holds up in experiment or particularly noisy eigentasks preclude any meaningful use of a given quantum reservoir for actual computation.
Fig.~\ref{fig:rec_time_steps_finite_lin_3_20_0.04} depicts the REC values for the same reservoirs as in Fig.~\ref{fig:rec_time_steps_infinite_3_0.04}, but obtained by using $S = 2^{20}$ measurement shots.
As expected, approximating expectation values via finite statistics leads to a reduced expressivity.
Only the four qubit system reaches values close to the infinite-shot case, since the number of readouts and the overall state space is much smaller than the number of injected linearly independent functions of the input.
The five and six qubit systems exhibit an increase of $C_{\text{T}}$ from the second to the third time step, but do not reach full saturation of their theoretically maximal REC during the three time steps included in this simulation.

Interestingly, the expressivity of the seven qubit system stays below the one of the six qubit system.
We attribute this to the exponential concentration of the output observables found for quantum systems \cite{xiong2025role, sannia2025exponential}.
In practice, it means that for an increasing qubit number, it becomes exponentially harder to extract meaningful information from the system, i.e., one has to perform exponentially more measurements to extract the same amount of information.
Here, the number of measurements are too few, such that the seven qubit systems is less expressive than the six qubit reservoir.
In App.~\ref{app:influence_of_number_of_measurement_shots}, we further discuss the influence of the shot number on the expressivity.

Owing to the finite shot measurements, we find the dissipative dynamics to reduce the REC starting from the fourth time step, where we do not include further inputs to the REC statistics.
As such, the reduced REC is expected and will not occur in conventional QRC, where new input is continuously injected, increasing the expressive capacity until limited by either the number of readouts or the combination of shot noise and dissipation.
Interestingly though, Fig.~\ref{fig:rec_time_steps_finite_lin_3_20_0.04} allows insight into a trend: larger reservoirs lose information more quickly than smaller versions with otherwise equal parameters.
This represents a major scalability issue for QRC with extreme scrambling reservoirs as it shows that the information storage time is inherently limited by the reservoir size.

\begin{figure}[h]
    \centering
    \includegraphics[scale=1]{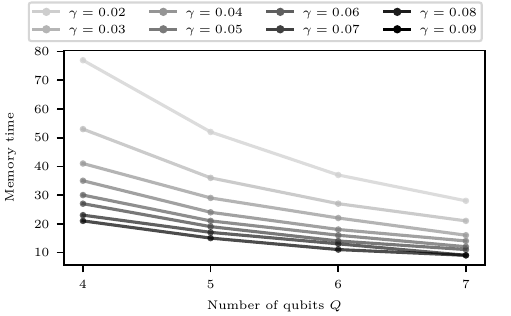}
    \caption{
        Memory time of a QRC, realized by a Haar random unitary, in dependence on the reservoir size $Q$.
        Gray scales represent different decay rates.
        Results are averaged over ten different realizations of the Haar random unitary.
        Shaded areas indicate the standard deviation.
        The memory time is defined to be the number of time steps, for which $C_{\text{T}} \geq 1.5$ holds.
    }
    \label{fig:memory_time_finite_lin_4_1_20}
\end{figure}

To further investigate this trade-off between the system size and the memory retention, we calculate the one-dimensional REC in dependence on the time steps using $S = 2^{20}$ measurement shots.
From this we extract the memory time $L_{\text{m}}$, which we define to be the number of time steps where the REC is $C_{\text{T}} \geq 1.5$ \footnote{The threshold of $C_{\text{T}} = 1.5$ is quite arbitrary, but motivated by the fact, that a capacity of $C_{\text{T}}=1$ could origin only from a constant, independent of the input. Hence, as a rule of thumb, if the capacity is falls below $C_{\text{T}}=1.5$ nearly all non-linear information is gone.}.
We choose this threshold as the constant function always ensures an REC of at least one, meaning that a reservoir with $C_{\text{T}} < 1.5$ is barely able to even represent a single nonlinear function of the input.
Results are depicted in Fig. \ref{fig:memory_time_finite_lin_4_1_20}.
Consistent with our previous observation, the memory time is lower for larger systems, indicating that the eigentasks of larger reservoirs fade away faster.
In agreement with the intuitive expectation, stronger dissipation further restricts the memory time.
Our analysis is solely based on extreme scrambling reservoirs, however a similar behavior of decreasing memory time with the system size has already been observed in Ref. \cite{xiong2025role}.

We can identify two regimes of reservoir sizes.
In the regime of too small reservoirs, the system is too small to capture all non-linear functions of the input.
In the regime of too large reservoirs, the system offers enough degrees of freedom for the injected information to unfold in the phase space, but the memory time reduces drastically, again posing a performance limit.
We thus propose the existence of a sweet spot between these two regimes that combines optimal memory with a phase space of sufficient size.
This thought is expanded in the discussion of Sec.~\ref{sec:optimal_reservoir_size}.


\begin{figure}[h]
    \centering
    \includegraphics[scale=1]{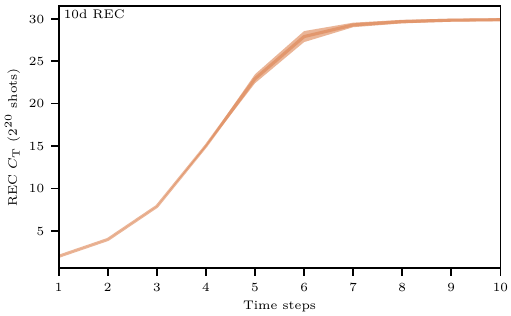}
    \caption{
        Resolvable expressive capacity for a $Q=5$ qubit reservoir using $2^{20}$ measurement shots for a decay rate of $\gamma = 0.04$.
        The input is injected with the linear encoding scheme on $Q_{\text{in}}=1$ input qubit.
        The reservoir dynamics are realized by a Haar random unitary and the results are averaged over ten realizations.
        Shaded areas show the standard deviation.
        For this simulation, the inputs of all ten time steps were included as independent variables in the REC analysis.
    }
    \label{fig:rec_time_steps_finite_single_lin_5_10_20}
\end{figure}

To illustrate how the REC develops over more than three time steps, we now approximate it for a $Q=5$ qubit reservoir with $Q_{\text{in}}=1$ input qubit over ten input cycles.
As mentioned above, the number of trajectories needed for the REC analysis scales exponentially with the number of layers, making the simulation a costly process.
We thus limit the number of inputs per layer to $N=3$, which matches exactly the expected number of linearly independent output functions the input encoding gives rise to.
While reducing the number of inputs per layer enables analysis for longer times, it also leads to a mild underestimation of the REC as discussed in \cite{hu2023tackling} that has to be kept in mind when discussing the results.
The REC for a decay rate of $\gamma = 0.04$ is depicted in Fig. \ref{fig:rec_time_steps_finite_single_lin_5_10_20}.
As predicted, the multidimensional expressive capacity grows in time due to new information injections at each time step, until (a) the upper bound of the state space size is reached, (b) the upper bound of the number of readouts is reached or (c) there is an equilibrium between the fading of past information and the injection of new information.
In the illustrated case for $\gamma = 0.04$, we find the REC to indeed come close to the maximum of $2^5 = 32$.
We attribute this difference to the noise and damping in the system as well as the underestimation of the REC due to limited input sequence samples.

Summarizing our findings in this section, we have numerically shown that the expressivity of a QRC grows with the number of time steps as expected.
However, this scaling has limitations: Firstly, the size of the reservoir determines how many degrees of freedom are accessible for storing non-linear functions of the input.
Secondly, the number of readouts trivially restricts the size of the output function space.
Thirdly, the interplay of dissipation, the exponential concentration of output observables and the approximation of expectation values with finitely many measurements inherently restricts the memory time of extreme scrambling reservoirs, hinting that small-sized QRC systems are more viable for classical time series processing.

\subsection{Input encoding} \label{sec:input_encoding}

Input encoding plays a vital role in QML \cite{schuld2021effect, govia2022nonlinear, mujal2021analytical}.
In the following, we want to investigate the dependence of the expressivity for two input encoding schemes and a variable number of input qubits.

\begin{figure}[h]
    \centering
    \includegraphics[scale=1]{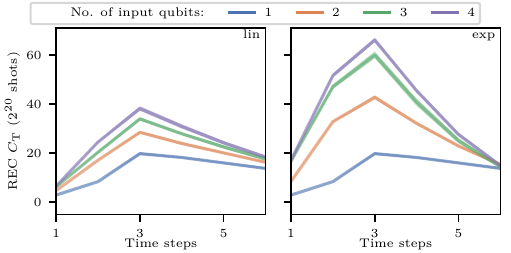}
    \caption{
        Three dimensional resolvable expressive capacity in dependence on the QRC cycles for a $Q=7$ qubit reservoir.
        The input is injected either with the linear (left) or the exponential (right) encoding scheme on a variable number of input qubits.
        The REC has been calculated from approximate expectation values using $2^{20}$ measurement shots.
        The reservoir dynamics are realized by a Haar random unitary and the results are averaged over ten realizations.
        Shaded areas indicate the standard deviation.
        For the first two time steps, the eigentasks are, of course, not three dimensional but one and two dimensional, respectively.
        For the simulation a decay rate of $\gamma = 0.04$ has been used.
    }
    \label{fig:rec_time_steps_input_qubits_finite_7_3_20_0.04}
\end{figure}

Fig. \ref{fig:rec_time_steps_input_qubits_finite_7_3_20_0.04} shows the REC values obtained from finite measurement shots in dependence on the number of input qubits $Q_{\text{in}}$, while the system size is constant at $Q=7$.
The case of only one input gate underpins our conclusion from Sec. \ref{sec:multivariate_fourier_decomposition} that increasing the size of the total reservoir by adding more qubits does not directly increase the expressivity.
At time step 3 the expressivity is limited to $C_{\text{T}} = 27$ due to the input encoding.
Although the seven qubit reservoir offers many more degrees of freedom, these are not used.
If the input encoding is improved, e.g. using an additional input qubit, the REC increases.
Furthermore, during the first three time steps the expressivity is only increased by injecting new inputs to the quantum system in a non-linear way.
After the third time step, the expressivity decreases due to the dissipation.
Non-linearity in QRC comes solely from the input encoding \cite{mujal2021analytical,govia2022nonlinear}, preventing QRC from composing non-linear dependencies on the input by time evolution.

Comparing the linear with the exponential encoding scheme shows that in the case of $Q_{\text{in}} = 1$, they are identical as predicted by the analytical expression in equations \eqref{eq:frequency_scaling_lin} and \eqref{eq:frequency_scaling_exp}.
In cases of $Q_{\text{in}} > 1$, the exponential encoding can help to inject information more efficiently into the reservoir.
By efficiently, we mean that more linearly independent functions of each input signal (in each time step) are written into the system.
We already observed in Sec. \ref{sec:reservoir_size_and_memory_time} that the linear encoding scheme does not saturate the state space size nor the number of readouts of the seven qubit system.
This is also the case for the exponential encoding, but it significantly increases the information content in the system.
One may be tempted to think that after this, the exponetial encoding is superior over the linear one.
However, as we will describe in Sec. \ref{sec:optimal_reservoir_size}, there are cases which benefit from switching from the exponential to the linear scheme.


To this point, we have shown that the size of the reservoir in QRC plays a very different role compared to RC with classical systems.
In classical RC, the reservoir nodes are coupled non-linearly, i.e., the information is processed in a non-linear way during the time evolution of the system.
Therefore, the output function space can be enriched by increasing the number of computational nodes.
In quantum mechanics, time evolution is inherently linear.
Once information is injected, it will only be linearly distributed between the degrees of freedom, the number of qubits merely controlling the amount of said degrees of freedom.

\subsection{Optimal reservoir size} \label{sec:optimal_reservoir_size}

Since the number of linearly independent multivariate functions grows exponentially with the number of processed time steps in QRC, the larger Hilbert space offered by larger reservoirs comes in favor when processing sequences with long coherence times.
However, as illustrated in Sec. \ref{sec:reservoir_size_and_memory_time}, larger reservoirs have a shorter intrinsic memory time.
So, while larger reservoirs might offer a large Hilbert space, before this space is used for computation, the reservoir forgets earlier inputs.
\begin{figure}[h]
    \centering
    \includegraphics[scale=0.96]{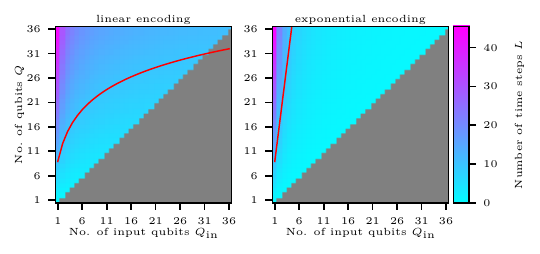}
    \caption{
        Theoretical number of time steps for which the encoded information (determined by the number of input qubits $Q_{\text{in}}$ and the encoding scheme) can be stored in all degrees of freedom of the reservoir of a certain size $Q$.
        The possible number of time steps for which information can be stored have been approximately calculated with Eqs. \eqref{eq:approx_time_step_scaling_lin} and \eqref{eq:approx_time_step_scaling_exp} using: $Q \approx \lceil \text{log}_4 [(2 Q_{\text{in}} + 1)^L] \rceil$ and $Q \approx \lceil \text{log}_4 [3^{Q_{\text{in}} L}] \rceil$.
        The grayed-out triangular region indicates $Q_{\text{in}} > Q$.
        Assuming, that the reservoir can store information for $L_{\text{m}} = 10$ time steps, the red line represents the boundary, beyond which is obstructive to further increase the system size, because it would decrease the memory time.
        Below the red curve, the reservoir would theoretically be able to store all information from the $L = 10$ time steps in its degrees of freedom.
    }
    \label{fig:n_qubits_limit_colormap}
\end{figure}
To visualize the interplay between the total system size, number of input qubits, and the processed time steps, we calculate how many time steps can be processed by the reservoir, without running out of degrees of freedom.
I.e., a certain system size gives rise to the available degrees of freedom and the number of input qubits and the encoding scheme determine how many degrees of freedom get occupied by linearly independent functions of each input signal.
We calculate the number of time steps after which all degrees of freedom are occupied in the ideal case of infinite shot statistics, thus providing an upper bound of meaningful reservoir size.
This can be determined by setting equations \eqref{eq:frequency_scaling_lin} and \eqref{eq:frequency_scaling_exp} equal to the number of degrees of freedom.
For the linear encoding this gives:
\begin{align}
    &&(2 Q_{\text{in}} + 1)^L &= 4^Q - 1 \nonumber \\
    &\Leftrightarrow &Q &= \lceil \text{log}_4 [(2 Q_{\text{in}} + 1)^L + 1] \rceil \label{eq:reservoir_scaling_lin} \\
    &&&\approx \lceil L \text{log}_4 [(2 Q_{\text{in}} + 1)] \rceil \nonumber \\
    &\rightarrow &L &= \frac{Q}{\lceil \text{log}_4 [(2 Q_{\text{in}} + 1)] \rceil} \label{eq:approx_time_step_scaling_lin}
\end{align}
and analogously for exponential encoding:
\begin{align}
    &&Q &= \lceil \text{log}_4 [3^{Q_{\text{in}}L} + 1] \rceil \label{eq:reservoir_scaling_exp} \\
    &\rightarrow &L &\approx \frac{Q}{\lceil \text{log}_4 [3^{Q_{\text{in}}}] \rceil} \label{eq:approx_time_step_scaling_exp} .
\end{align}
Fig. \ref{fig:n_qubits_limit_colormap} depicts the corresponding results, where the number of time steps that can be processed is depicted as color map in the dependence on the number of system and input qubits.
The grayed-out triangular part represents the unphysical situation of more input qubits than total system qubits.

For fixed input qubits, increasing the number of system qubits increases the available degrees of freedom.
Hence, the system is able to store more information and process longer sequences.
Vice versa, keeping the number of system qubits fixed and increasing the number of input qubits, the processible sequence length decreases.
This is due to the specific working principle of input encoding in quantum systems.
As elaborated in Sec. \ref{sec:multivariate_fourier_decomposition} injecting input in a redundant way increases the number of non-linear functions of this particular input, and the available degrees of freedom are occupied after a shorter sequence.

Information injection can further be tuned via the encoding scheme.
Exponential encoding is much more efficient in writing non-linear information into the system.
The same number of linearly independent functions of the input can be encoded with fewer encoding gates, making this encoding scheme in most situations beneficial, especially due to its minimal pre-processing overhead.
However, as this advantage comes at the cost of requiring a larger reservoir to store all information, there are situations where one should prefer the linear encoding as we will explain later on.

Despite exponential growth of the state space, the sensible reservoir size is limited in QRC by the fact that the memory time decreases with the system size.
Hence, there exists a boundary beyond which it is obstructive to further increase the system size, since additional degrees of freedom would not be used.
This boundary has to be determined for each setup individually, since it depends on the following parameters:
\begin{itemize}
    \item number of used input qubits and the input encoding scheme; this determines how many linearly independent functions are injected per input datum
    \item dissipation rate which directly influences the memory time
    \item number of measurement shots, which determines the approximation accuracy of the observable's expectation values.
\end{itemize}
This poses the important question about the optimal size for a QRC.
We propose the following scheme to determine it:
\begin{enumerate}
    \item Fix the number of input qubits $Q_{\text{in}}$ and the encoding scheme.
    \item For a given reservoir size, calculate the memory time $L_{\text{m}}$ using a fixed number of measurement shots.
    \item Insert $Q_{\text{in}}$ and $L_{\text{m}}$ into Eq. \eqref{eq:reservoir_scaling_lin} or \eqref{eq:reservoir_scaling_exp}.
    If the required number of qubits is smaller than the number of qubits used in the setup to determine $L_{\text{m}}$, reduce the number of qubits by one and repeat the process.
\end{enumerate}
Following the procedure up to 3, one has determined the maximal sensible reservoir size $Q_{\text{max}}$.
However, the optimal size $Q_{\text{opt}}$ is possibly even lower, since, due to dissipation and measurement noise, the REC might not saturate the number of readouts or the number of degrees of freedom.
As we saw in Sec. \ref{sec:input_encoding}, if the number of readouts scales with the system size, lowering the system size, will probably also lead to a lower attainable REC, even if the REC does not saturate the number of readouts nor the number of degrees of freedom. In such a case, $Q_{\text{max}}$ might be already the optimal reservoir size.
However, in case the number of readouts does not scale with the system size, one can follow the next steps, to determine the optimal reservoir size:
\begin{enumerate}
  \item[4.] Perform the multivariate REC analysis for a system size of $Q = Q_{\text{max}}$.
  \item[5.] Set $Q \rightarrow Q - 1$.
  \item[6.] Perform the multivariate REC analysis for a system size of $Q$.
\end{enumerate}
The optimal reservoir size is given by $Q_{\text{opt}}$ if $Q_{\text{opt}} - 1$ has a lower maximal attainable REC.

If one encounters the situation where the system size imposes a memory time which is not long enough for the task at hand (because the task has a correlation time which is longer than the memory time), one can reduce the number of system qubits and the number of input qubits, simultaneously.
The former gives rise for a longer memory time while reducing the available degrees of freedom.
The latter reduces the number of linearly independent functions per dimension / input injection step, thus, given the opportunity to store information from more time steps.
Instead of reducing the number of input qubits, one can also change from the exponential to the linear encoding scheme, which has also leads to a reduced input injection.
The proposed procedure is quite tedious and depends heavily on the employed system, but it ensures that one does not unnecessarily restrict the memory time by choosing a too large system.

To give at least a practical example, we consider a system possessing a memory time of $L_{\text{m}} = 10$.
With this, we calculate via Eq. \eqref{eq:reservoir_scaling_lin} and \eqref{eq:reservoir_scaling_exp} the required number of qubits.
The result as function of the input qubits is plotted as red line in Fig. \ref{fig:n_qubits_limit_colormap}.
We would like to point out that the assumption of a const. $L_{\text{m}}$ independent of $Q$ is a strong simplification.
Nevertheless, the example neatly shows that it is suboptimal to increase the system size beyond the red boundary.
Beyond this boundary, it is counterproductive to add more system qubits, as it comes at the cost of decreasing the memory time.

\section{Conclusion} \label{sec:conclusion}

In our work we formalized the notion of expressivity in quantum reservoir computing (QRC).
We analytically showed that the output of a QRC can be interpreted as a multi-dimensional Fourier series, where each input time step gives rise to an additional dimension of the Fourier series.
We support our analytical findings with numerical simulations by extending the REC analysis to multivariate eigentasks and use the multi-dimensional resolvable expressive capacity (REC) as our main  quantification of expressivity.
In an idealistic setup without dissipation, the multi-dimensional expressivity grows exponentially in the number of time steps, even if a linear encoding scheme is used.

We found that the indefinite growth of expressivity is practically hindered by three factors.
A $Q$ qubit system has $4^Q-1$ degrees of freedom, limiting the maximal processible number of linearly independent functions to the same limit.
However, in real QRC implementations the expressivity will be most likely limited by the number of observables, due to the obstacle of extracting $4^Q-1$ readouts from the system.
Even if the expressivity does not hit the first two upper bounds, it will plateau at the point where the fading of past information is equal to the injection of new information.
Furthermore, quantum systems are prone to the exponential concentration phenomenon, which effectively decreases the time for which information can be stored with increasing system size.
Together with the former, this implies a bound on the system size beyond which it is superfluous to add more qubits to the system, because (a) it would further restrict the memory time and (b) the additional degrees of freedom would not be used.
To tackle this problem, we give an explicit design procedure for determining the size of the reservoir.

Our results are central for the design of QRC implementations and highlight the interplay of reservoir size, input encoding and memory time.
As our used reservoirs are solely realized by maximally scrambling Haar random unitaries, future work could investigate the expressivity of practical reservoirs, like gate-based systems or the transverse-field Ising model.

\vspace{1cm}

\paragraph*{Acknowledgements}
The authors would like to thank Frederik Lohof for many useful discussions. This work has been supported by the Quantum Computing Initiative of the German Aerospace Center (DLR) via the Quantum Fellowship Program. We are grateful for funding from the German Research Foundation (DFG) and the Agence nationale de la recherche (ANR) via the German-French cooperation project \emph{PhotonicQRC} (DFG: Gi1121/6-1).

\vspace{1cm}

\newpage
\clearpage
\newpage
\bibliography{main.bib}

\clearpage
\onecolumngrid

\appendix

\section{Fourier series representation of output observables in QRC} \label{app:fourier_series_representation_of_qrc}

Complementary to the derivation of the Fourier series representation of output observables given in Sec. \ref{sec:multivariate_fourier_decomposition} we provide a more detailed version in this appendix.
We consider an input injection with single-qubit gates with encoding Hamiltonians $G_{(q)}$:
\begin{align}
    S(u) = \text{e}^{- \text{i} u G_{(0)}} \otimes \ldots \otimes \text{e}^{- \text{i} u G_{(Q - 1)}} .
\end{align}
Because all single-qubit gates commute, they can be simultaneously diagonalized with $G_{(q)} = V_{(q)}^{\phantom{\dagger}} \sigma_z V_{(q)}^\dagger$:
\begin{align}
    S(u) &= V_{(0)}^{\phantom{\dagger}} \text{e}^{- \text{i} \frac{u}{2} \sigma_z} V_{(0)}^\dagger \otimes \ldots \otimes V_{(Q - 1)}^{\phantom{\dagger}} \text{e}^{- \text{i} \frac{u}{2} \sigma_z} V_{(Q - 1)}^\dagger \nonumber \\
    &\coloneq V \text{exp} \left( - \text{i} \frac{u}{2} \sum_{q=0}^{Q - 1} \sigma_z^{(q)} \right) V^{\dagger} \nonumber \\
    &\coloneq V \text{e}^{- \text{i} u \Sigma} V^{\dagger} ,
\end{align}
where $\sigma_z^{(q)}$ is a multi-qubit operator which acts only via $\sigma_z$ on the $q$-th qubit:
\begin{align}
    \sigma_z^{(q)} = \underbrace{I \otimes \ldots \otimes \sigma_z}_{q \text{ times}} \otimes \ldots \otimes I .
\end{align}
$\Sigma$ is a $D \times D$ diagonal matrix with its eigenvalues $\lambda_d$ on the diagonal.
The system is initialized in a, generally, mixed state
\begin{align}
    \rho_0 = \sum_{i,j = 0}^{D - 1} \rho_{ij} \ket{i} \bra{j} .
\end{align}
$i$ and $j$ are written in binary.
The reservoir dynamics are described by a unitary $U_{\text{R}}$:
\begin{align}
    U_{\text{R}} = \sum_{i,j = 0}^{D - 1} w_{ij} \ket{i} \bra{j} .
\end{align}
After $L$ successive applications of the input gate and the reservoir gate, we find the state of the system to be:
\begin{align}
    \rho &= U_{\text{R}} S(u_{L - 1}) \cdots U_{\text{R}} S(u_0) \rho_0 S^\dagger (u_0) U_{\text{R}} \cdots S^\dagger (u_{L - 1}) U_{\text{R}} \nonumber \\
    &= \sum_{d_l^{\phantom{\prime}}, d_l^{\prime} \in [|1,D|]} \text{e}^{\text{i} (\lambda_{d_0^\prime} - \lambda_{d_0^{\phantom{\prime}}}) u_0} \cdots \text{e}^{\text{i} (\lambda_{d_{L - 1}^\prime} - \lambda_{d_{L - 1}^{\phantom{\prime}}}) u_{L - 1}} w_{i d_L^{\phantom{\prime}}} w_{d_L^{\phantom{\prime}} d_{L - 1}^{\phantom{\prime}}} \dots w_{d_2^{\phantom{\prime}} d_1^{\phantom{\prime}}} \rho_{d_1^{\phantom{\prime}} d_1^\prime} w_{d_2^\prime d_1^\prime}^* \dots w_{d_L^\prime d_{L - 1}^\prime}^* w_{j d_L^{\prime}}^* \ket{i} \bra{j} \nonumber \\
    &= \sum_{\bm{d}, \bm{d}^\prime \in [D]^L} \text{e}^{\text{i} (\bm{\lambda}_{\bm{d}^\prime} - \bm{\lambda}_{\bm{d}}) \bm{u}} w_{i d_L^{\phantom{\prime}}} w_{d_L^{\phantom{\prime}} d_{L - 1}^{\phantom{\prime}}} \dots w_{d_2^{\phantom{\prime}} d_1^{\phantom{\prime}}} \rho_{d_1^{\phantom{\prime}} d_1^\prime} w_{d_2^\prime d_1^\prime}^* \dots w_{d_L^\prime d_{L - 1}^\prime}^* w_{j d_L^{\prime}}^* \ket{i} \bra{j} \nonumber ,
\end{align}
where $[D]^L = \{ (d_1, \ldots, d_L)^{\text{T}} | d_l \in [|0, D - 1|] \}$.
As described in the main text, $P$ output observables $\{ O^{(p)} = \sum_{ij}^{D} O_{ij}^{(p)} \ket{i} \bra[j] \}_{p = 0, \ldots, P - 1}$ are used to extract information at the $L$-th time step:
\begin{align}
     \langle O^{(p)} \rangle (\bm{u}) & = \sum_{ij}^{D} \sum_{\bm{d}, \bm{d}^\prime \in [D]^L} \text{e}^{\text{i} (\bm{\lambda}_{\bm{d}^\prime} - \bm{\lambda}_{\bm{d}}) \bm{u}} w_{i d_L^{\phantom{\prime}}} w_{d_L^{\phantom{\prime}} d_{L - 1}^{\phantom{\prime}}} \dots w_{d_2^{\phantom{\prime}} d_1^{\phantom{\prime}}} \rho_{d_1^{\phantom{\prime}} d_1^\prime} w_{d_2^\prime d_1^\prime}^* \dots w_{d_L^\prime d_{L - 1}^\prime}^* w_{j d_L^{\prime}}^* O_{ji}^{(p)} \nonumber \\
    &= \sum_{\bm{\bm{\omega} \in \Omega}} \text{e}^{\text{i} \bm{\omega} \bm{u}} c_{\bm{\omega}} , \nonumber
\end{align}
with $\Omega$ and $c_{\bm{\omega}}$ from the main text.

\section{Multivariate REC analysis} \label{app:multivariate_rec_analysis}

Here, we give more information on the calculation of the multivariate eigentasks.
A very detailed description of the univariate case can be found in Ref. \cite{hu2023tackling}.
Considered is a physical (quantum) system, which takes $F$ inputs $\bm{u} \in \mathbb{R}^F$.
The system processes the inputs and for information extraction measurements are performed on $R$ degrees of freedom yielding the measured features $\bar{X}_r (\bm{u}), r \in \{ 0, \ldots, R-1\}$ by averaging over $S$ shots.
The goal is to approximate a target function $f(\bm{u})$ by using a linear combination of the measured features
\begin{align} \label{eq:measured_features_linear_combination}
    f_{\bm{\eta}} (\bm{u}) = \sum_{r=0}^{R-1} \eta_p \bar{X}_{r} (\bm{u}) \quad ,
\end{align}
where $\bm{\eta} = (\eta_0, \ldots, \eta_{R-1})^{\text{T}}$ are the trainable weights.
Comparing this with the QRC procedure, the measured features are the expectation values of observables $\bar{X}_r (\bm{u}) = \langle O^{(r)} \rangle$ and Eq. \eqref{eq:measured_features_linear_combination} is exactly the linear layer from Eq. \eqref{eq:prediciton_linear_combination}.
The REC analysis answers the question, which orthogonal functions $f(\bm{u})$ can be approximated by Eq. \eqref{eq:measured_features_linear_combination} with minimal error (meaning that the normalized mean-squared accuracy is maximized) \cite{hu2023tackling}.
Then, output of the REC calculation is the resolvable expressive capacity (REC) $C_{\text{T}}$, quantifying how many orthogonal functions can be approximated by Eq. \eqref{eq:measured_features_linear_combination} and the eigentasks $\{y^{(r)} (\bm{u})\}_{r = \{0, \ldots, R-1\}}$ itself, representing the orthogonal functions which saturate the normalized mean-squared accuracy.

Applying this to QRC, $\bm{u}$ corresponds to the injected time series $\bm{u} = (u_1, \ldots, u_{L})^{\text{T}}$.
However, since the numerical calculations are quite expansive, we use only the first $\tilde{l}$ inputs for the eigentask construction.
In particular, we inject $u_1, \ldots, u_{\tilde{l}} \in [- \pi, \pi]$ (each $N$ values), followed by an injection of the remaining inputs $u_{\tilde{l}}, \ldots, u_{L}$ which are random, but fixed for all values of $u_1, \ldots, u_{\tilde{l}}$.
With this, the eigentasks become functions of the first $\tilde{l}$ inputs: $\{ y^{(r)} (\bm{u}_{\tilde{l}}) \}_{r=\{0, \ldots, R-1\}}$, where $\bm{u}_{\tilde{l}} = (u_1, \ldots, u_{\tilde{l}})$.
Since we use single-qubit rotations as input gates, the natural interval of the inputs is $[- \pi, \pi]$, i.e., all first $\tilde{l}$ inputs (which are the independent variables of the eigentasks) lie in this interval.
Hence, we choose an $N$-step discretization of the interval $[- \pi, \pi]$: $U_N \coloneq \{- \pi + \frac{2 \pi k}{N - 1}| k = 0, \ldots, N - 1\}$, which serves as input values for each $u_1, \ldots, u_{\tilde{l}}$.
In the following set we collect all input tuples for the first $\tilde{l}$ time steps which are used to obtain the output of the QRC which is then used to calculate the eigentasks:
\begin{align}
    S &= U_N \times \ldots \times U_N \\
    &= \{ (u_1, \ldots, u_{\tilde{l}}) | u_j \in U_N \, \forall j = 1, \ldots, \tilde{l} \} \quad .
\end{align}
The remaining inputs (which are not treated as independent variables of the eigentasks) are randomly sampled from $[- \pi, \pi]$ but fixed for all input tuples of $S$.

The measured features are then the expectation values of observables after time step $L$.
All these features will be collected in a single regression matrix, where measured features are squeezed along the inputs of all dimensions:
\begin{align}
    \begin{pmatrix}
        \bar{X}_0 (u_1^{(0)}, \ldots, u_{\tilde{l} - 1}^{(0)}, u_{\tilde{l}}^{(0)}) & \dots & \bar{X}_{R-1} (u_1^{(0)}, \ldots, u_{\tilde{l} - 1}^{(0)}, u_{\tilde{l}}^{(0)}) \\
        \bar{X}_0 (u_1^{(0)}, \ldots, u_{\tilde{l} - 1}^{(0)}, u_{\tilde{l}}^{(1)}) & \dots & \bar{X}_{R-1} (u_1^{(0)}, \ldots, u_{\tilde{l} - 1}^{(0)}, u_{\tilde{l}}^{(1)}) \\
        \vdots & & \vdots\\
        \bar{X}_0 (u_1^{(0)}, \ldots, u_{\tilde{l} - 1}^{(0)}, u_{\tilde{l}}^{(N - 1)}) & \dots & \bar{X}_{R-1} (u_1^{(0)}, \ldots, u_{\tilde{l} - 1}^{(0)}, u_{\tilde{l}}^{(N - 1)}) \\
        \bar{X}_0 (u_1^{(0)}, \ldots, u_{\tilde{l} - 1}^{(1)}, u_{\tilde{l}}^{(0)}) & \dots & \bar{X}_{R-1} (u_1^{(0)}, \ldots, u_{\tilde{l} - 1}^{(1)}, u_{\tilde{l}}^{(0)}) \\
        \vdots & & \vdots \\
        \bar{X}_0 (u_1^{(N - 1)}, \ldots, u_{\tilde{l} - 1}^{(N - 1)}, u_{\tilde{l}}^{(N - 1)}) & \dots & \bar{X}_{R-1} (u_1^{(N - 1)}, \ldots, u_{\tilde{l} - 1}^{(N - 1)}, u_{\tilde{l}}^{(N - 1)})
    \end{pmatrix}
\end{align}
Using this single regression matrix, the multivariate REC calculation can be done as in the univariate case following the recipes in Ref. \cite{hu2023tackling}.
At the end, one just has to made sure that the eigentasks are unsqueezed accordingly.

\section{Influence of number of measurement shots} \label{app:influence_of_number_of_measurement_shots}

\begin{figure}[h]
    \centering
    \includegraphics[scale=1.4]{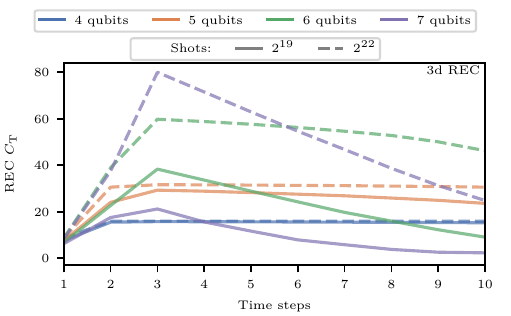}
    \caption{
        Three dimensional resolvable expressive capacity in dependence on the QRC cycles for different numbers of measurement shots.
        The input is injected either with the linear encoding scheme on $Q_{\text{in}} = 4$ input qubits.
        The reservoir dynamics are realized by a Haar random unitary and the results are averaged over ten realizations.
        We omit to plot the standard deviations.
        For the first two time steps, the eigentasks are, of course, not three dimensional but one and two dimensional, respectively.
        For the simulation a decay rate of $\gamma = 0.04$ has been used.
    }
    \label{fig:rec_time_steps_shots_qubits_lin_3_0.04}
\end{figure}

\begin{figure}[h]
    \centering
    \includegraphics[scale=1.4]{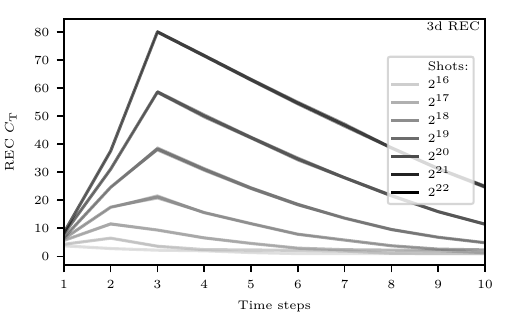}
    \caption{
        Three dimensional resolvable expressive capacity in dependence on the QRC cycles for a $Q = 7$ qubit system using different numbers of measurement shots.
        The input is injected either with the linear encoding scheme on $Q_{\text{in}} = 4$ input qubits.
        The reservoir dynamics are realized by a Haar random unitary and the results are averaged over ten realizations.
        Shaded areas indicate the standard deviation (which are very small).
        For the first two time steps, the eigentasks are, of course, not three dimensional but one and two dimensional, respectively.
        For the simulation a decay rate of $\gamma = 0.04$ has been used.
    }
    \label{fig:rec_time_steps_shots_7_4_lin_3_0.04}
\end{figure}

In this section, we briefly discuss the dependence of the resolvable expressive capacity (REC) on the number of measurement shots.
Figs. \ref{fig:rec_time_steps_shots_qubits_lin_3_0.04} and \ref{fig:rec_time_steps_shots_7_4_lin_3_0.04} depict the REC values for the same setup as used for Fig. \ref{fig:rec_time_steps_finite_lin_3_20_0.04}, except that expectation values are approximated using different number of measurement shots.
Since the expectation values of extreme scrambling systems exponentially concentrate around an input independent value with increasing reservoir size \cite{xiong2025role}, one needs exponentially many measurement shots to extract information from larger systems.
This caused the smaller REC for $Q=7$ compared to the $Q=6$ system in Fig. \ref{fig:rec_time_steps_finite_lin_3_20_0.04}.
Now, Fig. \ref{fig:rec_time_steps_shots_qubits_lin_3_0.04} shows that the $Q=7$ system can have a larger expressivity when using an appropriate number of shots.
However, it is still interesting that the seven qubit system has a larger REC that the six qubit system for $2^{22}$ shots, although the saturation bound ($2^6 = 64$ readouts) of the latter is not reached.
We attribute this to the larger readout dimension of the seven qubit system, from which more (noisy) information can be extracted.

\end{document}